\documentclass[proof]{WileyASNA-v1}

\articletype{Article Type}%

\received{xxxxxxxx}
\revised{xxxxxxxx}
\accepted{xxxxxxxx}

\begin{document}

\title{Photometric Search for Exomoons by using
Convolutional Neural Networks}

\author[1]{Lukas Weghs*}

\authormark{Lukas Weghs}

\address[1]{\orgname{Municipal Gymnasium Thomaeum}, \orgaddress{\state{North Rhine-Westphalia}, \country{Germany}}}

\corres{*Lukas Weghs, \email{L.Weghs@t-online.de}}

\abstract{Until now, there is no confirmed moon beyond our solar system (exomoon). Exomoons offer us new possibly habitable places which might also be outside the classical habitable zone. But until now, the search for exomoons needs much computational power because classical statistical methods are employed. It is shown that exomoon signatures can be found by using deep learning and Convolutional Neural Networks (CNNs), respectively, trained with synthetic light curves combined with real light curves with no transits. It is found that CNNs trained by combined synthetic and observed light curves may be used to find moons bigger or equal than roughly 2-3 earth radii in the Kepler data set or comparable data sets. Using neural networks in future missions like Planetary Transits and Oscillation of stars (PLATO) might enable the detection of exomoons.}



\maketitle

\section{Introduction}
In the past, machine learning was already applied in the search for exoplanets; \cite{mccauliff2015automatic} and \cite{mislis2016sidra} used for example random forest classification algorithms \citep{breiman2001random} to identify Transit Crossing Events (TCEs) in Kepler data. Besides, Convolutional Neural Networks \citep{lecun1989handwritten, lecun1998gradient} have successfully detected or re-identified exoplanet (candidates), respectively, by using CNNs in various different surveys like in the Next Generation Transit Survey \citep{chaushev2019classifying, wheatley2013next}, in TESS data \citep{osborn2020rapid} and in the K2 mission \citep{dattilo2019identifying}.\\
Exomoons are a new branch of exoplanetary research; these moons beyond our solar system might expand the classical understanding of habitability too. \cite{heller2020astrophysical} and \cite{tjoa2020subsurface} showed that for example through tidal heating subsurface oceans could be heated up, so that there could be liquid water even outside the classical habitable zone.\\ A survey that searched in the last years for exomoons was the Hunt for Exomoons by using Kepler (HEK) as described by \cite{kipping2012hunt}. But until now, analyzing and modeling the exomoon candidates from photometric time series is very computationally intensive (around 33,000 hours per target averagely) because researchers use classical numerical methods \citep{kipping2015hunt}. Convolutional neural networks might offer a possibility to speed up the process of identifying possible exomoon candidates and to establish a hierarchy of these candidates to enable analyzing these candidates in more detail that are most likely exomoons before the other candidates. The use of neural networks in the search for exomoons is a very new approach; it was not until the summer of 2020 that scientists first published a successful attempt to use a CNN to detect exomoons in simulated data \citep{ExomoonCNN}. But they did not apply their network to real observational data. I applied my network to a combination of observed Kepler data and simulated light curves, thereby coping with the problem that no exomoons are confirmed until now. Using trained CNNs might significantly speed up the search for exomoons for example in the PLAnetary Transits and Oscillations (Plato) mission \citep{rauer2014plato}\.\\
In §\ref{sectionSimulations}, I will describe my simulations especially the choice of parameters and the used Kepler light curves. The used network and preprocessing as well as the test methods will be described in §\ref{sectionCNN} and my results will be shown in §\ref{sectionResults}.

\section{Simulations} \label{sectionSimulations}
\subsection{Orbital configurations in the simulations}
To limit the number of free parameters and the size of the training set, I used in this project a Sun-like star as central star of the planetary system (see Table \ref{tab:parametersStellar}). In addition, I only calculated stable exomoon orbits around the planets. Satellites cannot be located within the distance from their planet as stipulated by the Roche limit, which is defined by the following equation \citep{rocheLimit}\footnote{This is the equation for liquid satellites. Solid satellites can have a smaller orbit than liquid ones. Use of this formula ensures that all satellites (which are spherical in shape) are outside the Roche limit.}
\begin{equation}
a_{min}\backsimeq2.44 r_p\left(\frac{\rho_p}{\rho_m}\right)^{1/3}
\end{equation}
Here \(\rho_m \) and \(\rho_p \)  are the mean densities of the moon and the planet, respectively, and  \(r_p \) is the radius of the planet. If the orbit of the satellites were smaller than this limit, they would be torn apart by the tidal force. On the other hand, exomoons must be within the Hill radius in order to be able to orbit a planet stably, as determined by  \cite{Rodenbeck2019}.
\begin{equation}
R_{Hill} = a_p \left (\frac{m_p}{3m_s} \right)^{\frac{1}{3}}
\end{equation}
Where \(R_ {Hill} \) is the Hill radius, \(m_p \) is the mass of the planet, \(m_ s \) is the mass of the star and \(a_p \) is the semi-major axis of the planet’s orbit. In my simulations, I assumed that the moon’s orbit is ``retrograde'': in retrograde orbits, the orbit is stable up to a fraction of the Hill radius \(\eta \) of one. Prograde moons only have a stable orbit at  \(\eta\leq0,5\) \citep{domingos2006stable}, so I did not take these moons into consideration in the penultimate simulations.
Since I only defined the limits of the orbital period and the actual value within these limits is chosen randomly, the semi-major axis of the exoplanet’s orbit as well as that of the moon was not available at the beginning and was calculated according to Kepler’s third law by assuming a Jupiter-like mean density for the planet and an Earth-like mean density for the moon.\footnote{{As shown by \cite{seager2007mass}, it can be assumed that exoplanets that are significantly heavier than Earth, are not rocky or consist of iron and other minerals.}} The planetary and lunar radii vary within a certain range (see Table \ref {tab:parametersPlanet} and Table \ref{tab:parametersMoon}). The mass of the planet or moon is determined by assuming a spherical planet with  \(V = \frac{4}{3}\pi r^3\) and rearranging the formula for the mean density  \(\varrho = \frac{m}{V} \Leftrightarrow m = V \varrho \) to
\begin{table}[t]
 \caption{Main planetary parameters assumed for the last simulations}
\centering
\begin{tabular}{ll}

\hline
\hline
    Parameter&Values \\
     \hline
     Impact \(\vartheta_p\) & 0-1 \\
     Period [days] \(P_p\)& 10-70 \\
     Mass [Earth Mass]  \(m_{p}\)& \(\frac{4}{3}\pi\cdot r_p^3\)\\
     Radius [Earth Radii] \(r_p\) & 9-13\\
     mean Density [\(\frac{kg}{m^3}\)]& 1.326\\
     inclination \(i_p\) &\( cos^{-1}(\frac{r_s}{b_p})\)\\
     &\(<i_p<\frac{\pi}{2}\)\\
     \hline
\end{tabular}
\label{tab:parametersPlanet}
\end{table}

\begin{table}[t]
\caption{Stellar parameters used for the last simulations}
\centering
\begin{tabular}{ll}
\hline
\hline
    Parameter&Values \\
     \hline
     Radius [Solar Radii] \(r_s\)& 1\\
     Mass [Solar Masses] \(m_s\) & 1\\
     Limb-darkening \(q_1\) &  0.6\\
     Limb-darkening \(q_2\) &  0.4 \\
    \hline
          \end{tabular}
\label{tab:parametersStellar}
\end{table}
\begin{table}[t]
 \caption{Main parameters of the exomoon and the planet-moon system in the last simulations}
\centering
\begin{tabular}{ll}
\hline
\hline
    Parameter&Values \\
     \hline
     Mass \(m_m\) [Earth Masses]&1.00-64.17\\
     Radius \(r_m\) [Earth Radii]& 2-4\\
     Hill radius \(R_{Hill}\) [au]&0.0078-0.1651\\
     Roche limit [au]&0.0006-0.0008\\
     Semi-major axis \(a_m\)[au]&0.0021-0.0078\\
    Period \(P_m\) [days]& 1.5-6\\
    Mean density \(\frac{kg}{m^3}\)& 5.513\\
    Inclination \(i_m\)& \(cos^{-1}(\frac{r_s-cos(i_p)*b_p}{b_m})\)\\
    &\(<i_m<\frac{\pi}{2}\)\\
    Number of moons&1-5\\
     \hline
     \end{tabular}
\label{tab:parametersMoon}
\end{table}
\begin{equation}
m = \frac{4}{3}*\pi*r^3*\varrho.
\end{equation}The planetary mass is therefore between about 0.6 and 1.6 Jupiter masses. This could increase the probability of detecting an exomoon, since the moon’s mass is smaller than the planet’s mass and consequently the radius of the moon is smaller than the planet’s radius, as explained above; this is why a large central planet was used. Radii smaller than the Earth’s radii are difficult to detect in photometric Kepler data, so the Earth’s radius was used as a lower limit for the radius of the moon. An upper limit to planetary mass derives from the fact that hydrogen begins to fuse when the gravitational mass is greater than about 13 Jupiter masses \citep{perryman2018exoplanet}; in addition, another study states that planets can have a maximum mass of 10 Jupiter masses, but masses greater than 5 Jupiter masses are very rare \citep{schlaufman2018evidence}.
The relationship between the period \(P \) and the semi-major axis  \(a \) is again defined by Kepler's third law. The calculation of the semi-major axis is carried out for the moon as well as for the planet. I set the planetary impact parameter to 0 in earlier simulations: this causes the planet to orbit its star in the plane of the observer (and the star). The time of the first transit in these earlier simulations was between 0 and 2 days (see Table \ref {tab:parametersPlanet} for the current simulations). An overview of further parameters used can be found in the appendix (see tables in \ref {otherParameters}). I also experimented with other parameters and combinations, but it emerged that no transit signatures caused by the exomoon were then visible, and/or the semi-major axis of the moon’s orbit was outside the Hill radius.

\subsection{Advanced orbital configurations in the simulations}
\begin{figure}
\includegraphics[width = 8cm]{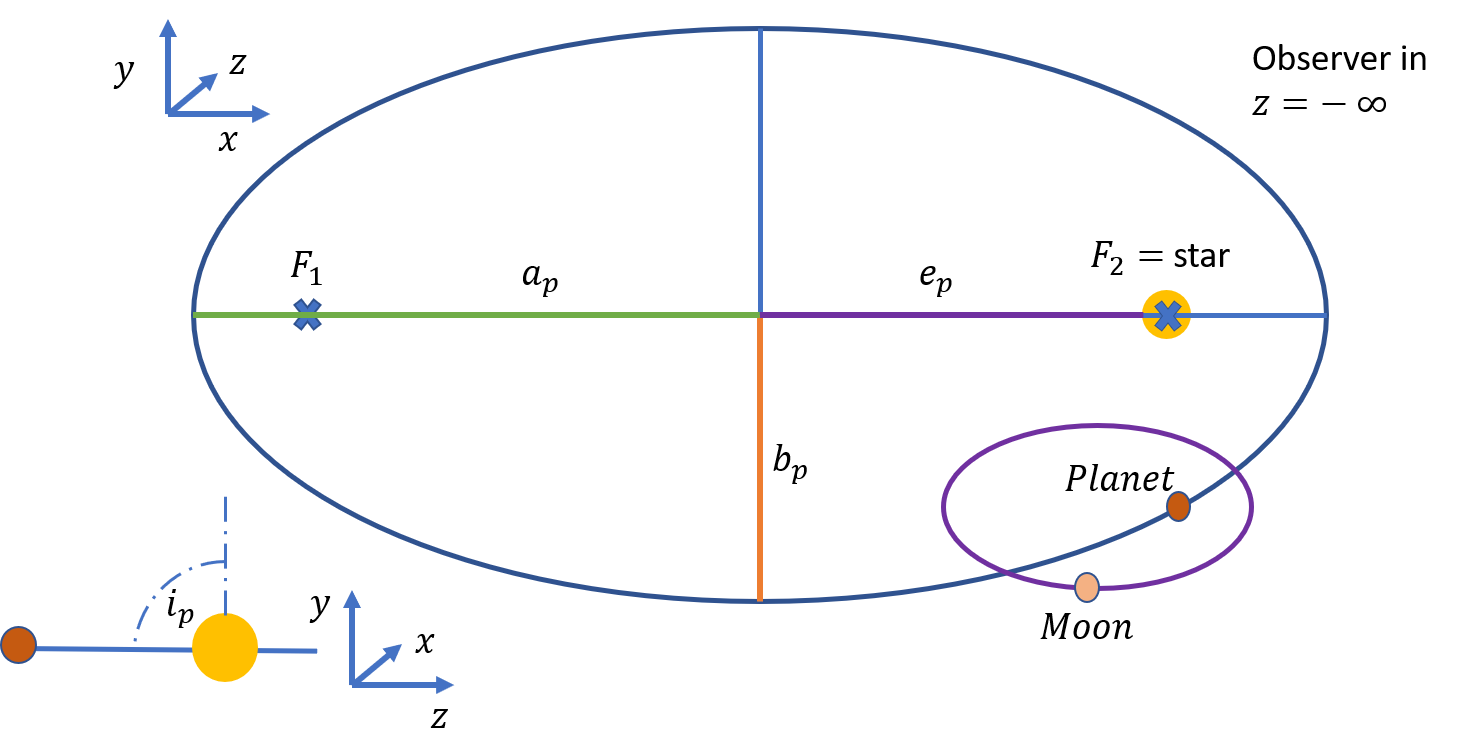}
\caption{The main parameters of the planetary orbit are delineated here. The parameters for the exomoon are designated in this project in the same way as those of the planet, with the difference that the planets are marked with a ``p'' and the moons with an ``m''. The semimajor axis of the exoplanet and -moon are assumed to be orthogonal to the line of sight between the observer and the star.}
\label{fig:OvervieworbitaleParameter}
\end{figure}

In order to be able to vary even more parameters in my simulations, I added the change in the inclination of the moon and the planet in the last simulation, which required several systems of equations to be set up. In these simulations I additionally varied the eccentricity of the planetary and lunar orbit as well as the inclination of the two systems and I optimized the determination of the time of the lunar transit. I also simulated both prograde and retrograde moons\footnote{It was found by \cite{perets2014formation} that during for example moon-moon-interactions possibly temporarily retrograde or even stable moons can be assumed.} in these data sets. Furthermore, I specified that 10\% of the systems with a moon should have multiple moons, and I simulated up to five different lunar transits near a planetary transit in these systems. The aim of simula\-ting several moons with different parameters is to reflect the situation in our solar system; however, at the same time I decided to simulate only a maximum of five lunar signatures for a planetary transit (and only in one tenth of the cases). This is because we are assuming very large moons here, so the existence of several such moons can be considered to be very unlikely. This was not something I wanted to rule out entirely either, however. For all the following equations, I assumed the situation shown in Fig. \ref{fig:OvervieworbitaleParameter}.
\subsubsection{Establishing the time of the exomoon transit}
To prevent falsifying the training set  of the CNN by exomoons which show no transit signal, I used geometric considerations to set up a collection of formulas,\footnote{The literature does not yet offer suitable formulas for this application.} which enable me to determine in the general case (of an eccentric orbit of the planet) when no transit occurs, thereby excluding these cases from the simulation in advance. By means of equivalent transformations of the center equation of an ellipse, I obtained the following functional formula.
\begin{equation}
\begin{split}
\frac{x^2}{a^2}+\frac{y^2}{b^2}=1
\Leftrightarrow y = \pm\frac{b}{a}\sqrt{a^2-x^2}
\end{split}
\end{equation}
Since in this system of equations only \(a\) was fixed in advance, I determined \(b\) from the numerical eccentricity \(\varepsilon\), which I also established using the formula  \(\varepsilon = \frac{e}{a}\), where \(e\) represents the linear eccentricity with \(e = \sqrt{a^2-b^2}\) . This resulted in the following formula \(b = \sqrt{a^2-\varepsilon^2a^2}\). For the sake of clarity, I show the most important of these parameters again in Fig. \ref {fig:OvervieworbitaleParameter}.
I reasoned that a lunar transit does not occur when the moon appears to be directly in front of/behind the planet. This is therefore the case in the interval \(I = [e-r_p; e+r_p]\). First I established the following formula, which enables me to calculate the time period in which the center of the moon is behind/in front of the planet, which of course depends on the radius of the planet \(r_p\).
\begin{equation}
\begin{split}
t_m =  \Big\{\int_{e_m-r_p}^{e_m+r_p}\frac{b_m}{a_m}\sqrt{a_m^2-x^2}\,dx \\
 - \frac{b_m}{a_m}\sqrt{a_m^2-(e_m-r_p)^2} \frac{r_p}{2}\\
 - \frac{b_m}{a_m}\sqrt{a_m^2-(e_m+r_p)^2} \frac{r_p}{2}  \Big\} \\
 \frac{P_m}{a_m b_m\pi}\\
\end{split}
\end{equation}
This results in the interval \(I_1 = [t_{0,m}-\frac{t_m}{2}; t_{0,m} +\frac{t_m}{2}]\), in which no transits occur for prograde or retrograde orbits of the moon where \(t_{0,m}\) is the time when the moon is the first time in front of its host. This interval indicates the time span in which the moon is in front of the planet. The time after which the moon is again behind the planet depends on the time it takes the moon to cover this distance in its orbit. This time \(t_1\) for a prograde moon is given by:
\begin{equation}
\begin{split}
&t_1  =  \Big\{2\int_{e_m+r_p}^{a}\frac{b_m}{a_m}\sqrt{a_m^2-x^2}\,dx \\
& +\frac{r_p b_m}{a_m}\sqrt{a_m^2-(e_m+r_p)^2}\Big\}\frac{P_m}{a_m b_m\pi}
\end{split}
\end{equation}
For retrograde moons, this period  \(t_2\),  after which the moon is again behind the planet, is given by the following formula:
\begin{equation}
\begin{split}
&t_2  = \Big\{2\int_{-a_m}^{e_m-r_p}\frac{b_m}{a_m}\sqrt{a_m^2-x^2}\,dx \\
& +\frac{r_p b_m}{a_m}\sqrt{a_m^2-(e_m-r_p)^2}\Big\}\frac{P_m}{a_m b_m\pi}
\end{split}
\end{equation}
So the second interval for prograde moons in which no transit is visible is \(I_2 = [t_{0,m}+\frac{t_m}{2}+t_1; t_{0,m}+\frac{3t_m}{2}+t_1]\) and for retrograde moons  \(I_2 = [t_{0,m}+\frac{t_m}{2}+t_2; t_{0,m}+\frac{3t_m}{2}+t_2]\). By assuming that the moon must not be in front of or behind the planet when this planet causes a transit, it is possible to establish \(t_{0,m} = t_{0,p}\).

\subsubsection{Inclination of the planetary orbit}
In the earlier simulations, the selected inclination \(i_m\) of the planetary orbit was exactly \(\frac{\pi}{2}\), which meant that the impact parameter did not vary during the planet’s transits, but this also has a major impact on the shape of a transit. In order to be able to introduce greater variance into my simulations here, I established a formula that allowed me to calculate the minimum inclination of the exoplanet. This minimum inclination is when the impact parameter is \( \vartheta_p = 1\).\footnote{In order to avoid confusion with the semi-minor axis, I deviate from the convention here in that I do not refer to the impact parameter here as \(b\).} I established the following formula for the impact parameter, set it as being equal to 1 and converted it according to the inclination.
\begin{equation}
\begin{split}
\vartheta_p = \frac{b_p  cos(i_{min,p})}{r_s} = 1\Leftrightarrow  i_{min, p} = cos^{-1}(\frac{r_s}{b_p})
\end{split}
\end{equation}

\subsubsection{Inclination of the lunar orbit}
In order for the moon to cause a transit in any case, it must pass in front of the star at all positions of its orbit (or at least at the position where it is at the time of the (theoretical) transit). This is the case when the moon has an inclination of \(\frac{\pi}{2}\), again measured relative to the plane perpendicular to the observer’s line of sight. But since this would again constitute a limitation of the parameter of space, I also varied this parameter in order to add even greater variability to my simulations. In order for the moon to cause a transit in any case (except in the two intervals \(I_1\) and \(I_2\)), however, I had to determine a minimum inclination \(i_{min,m}\). Again, on a geometric basis, I considered how the situation could be expressed in a formula, and I came up with the following formula which I transformed.
\begin{equation}
\begin{split}
b_m * cos(i_{min, m}) = r_s - cos(i_p) b_p\\
\Leftrightarrow i_{min,m} = cos^{-1}(\frac{r_s-cos(i_p) b_p}{b_m})
\end{split}
\end{equation}

\subsection{Simulation of the light curves}\label{simulation}
In general, for each set of parameters a light curve containing exomoon signatures and a light curve containing only the exoplanet signatures are calculated using the python-module planetplanet by \cite {luger2017planet}. Each of the calculated light curves contains 60,000 simulated photometric measurements between the Barycentric Julian dates 2454953 and 2456433, whereby an offset of 2454833 is subtracted in the Kepler data. This is based on Kepler’s observation time. This period and the 60,000 data points it contains result in a cadence of approximately 35.52 minutes: this is similar to the Kepler Long Cadence objects, which have a cadence of 29.4 minutes and each exposure consist of 270 sub-exposures \citep{perryman2018exoplanet}. The photometric accuracies of Kepler for certain targets are 30-40 ppm. In order to simulate this noise in the synthetic light curves, normal distributed noise with a standard deviation of 25\ ppm was added to the non-noisy transit signal\footnote{Instead of 30 ppm or 40 ppm as in the original Kepler data \citep {perryman2018exoplanet}, I used 25 ppm to make my results more comparable to  \cite{ExomoonCNN}.}   and the arithmetic mean of the light curve was set to one, which represents the normal brightness of the star. The light curves are randomly classified as part of the training or validation data set, with the first data set containing 90\% of the light curves and the second 10\% of the total simulated light curves. The entire training data set consists of 20,000 light curves – 10,000 that contain an exomoon and 10,000 that do not. The data set was calculated using distributed computing on the TU Berlin’s HPC cluster and is around 58 gigabytes in size.\\
In a further development step, I inserted the simulated light curves with the clean signature into original Kepler light curves so that the neural network can also be trained directly on the systemic noise of Kepler (instrumental errors) and does not recognize this noise as an exomoon (if the CNN did recognize the noise as an exomoon, it would be a false positive (FP)). The light curves into which I inserted the transit signals of the planets and the moons are referred to in the following as integration light curves, because here the transit signals are integrated in the light curves. For this purpose, I created an initial list of 72,507 stars whose magnitude is greater than 12 in the optical band (V-Band) \citep{van2016kepler} and which lie in Kepler’s field of view. For each of these stars, I then downloaded the light curve from NASA’s Mikulski Archive for Space Telescopes (MAST) and calculated the standard deviation for each of these light curves. If this deviation was less than 300 ppm, the light curve was selected, giving me 5,642 light curves. Exomoons and/or planets were then inserted into these light curves as described above. To increase the dataset, these Kepler light curves have been used several times adding each time artificial signal of  different configurations.\\
\begin{figure}[t]
\centering
 \includegraphics[width=8cm]{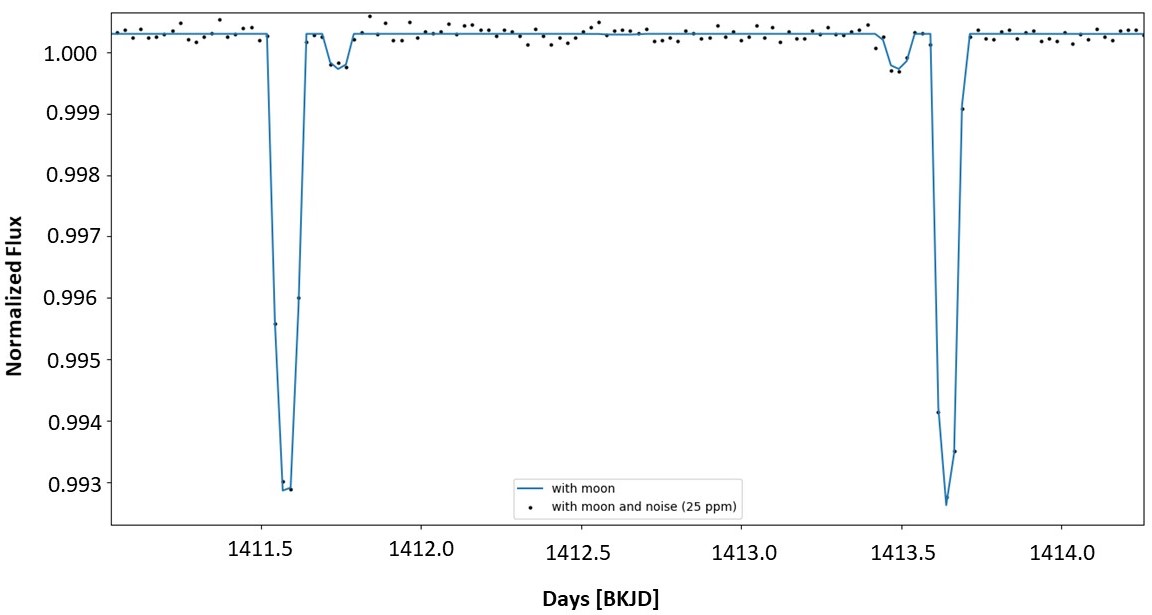}
 \caption{This graph shows the simulated light curve (with normally distributed noise) of a planet (\(P_p = 2 d\), \(R_p = 11 R_{\oplus}\)) by an exomoon (\(P_m = 0.4 d\), \(R_m = 3R_{\oplus}\)). The decrease in brightness to the right and left of the transits is caused by an exomoon. The neural network is trained to identify such transits of exomoons.}
 \label{fig:SimulatedExomoon}
\end{figure}

\subsubsection{Improvement of the selection procedure for the integration light curves}
In order to further differentiate the integration light curves, I calculated different noise levels that may be present in the individual light curves for my last simulations, thereby establishing four categories. In the first category there ought to be a relatively good SNR: here I set the limit at \(5\sigma \) transit depth of the smallest moon in this category. This resulted in a maximum noise of approx. 67 ppm. While I inserted a lunar transit in the first category, I kept the noise level the same in the second category, but I did not insert a lunar transit here; this means that the neural network in training can later exclude moons with a high probability in these light curves. In the third category I placed the integration light curves in which I only wanted to inject a planetary transit but which have a worse SNR. In my case, the planetary transit depth was again to produce at least \(5\sigma\) for the smallest planet. This resulted in a standard deviation of the integration light curve of between 67 ppm and 1316.6 ppm. In the last category I placed those integration light curves whose noise was between 67 ppm and 336 ppm. I injected a lunar transit here too, but with a poorer noise level; the transit depth for these light curves for  \(2R_\oplus\) moons is only between  \(1\sigma\) and \(5 \sigma\).

\section{Development, training and evaluation of the neural network} \label{sectionCNN}
The Convolutional Neural Network (CNN) was developed using the Keras Applied Programming Interface (API) by using Tensorflow as a backend \citep{tensorflow}.
\subsection{Architecture} \label{architecture}
The use of a CNN architecture to detect signals from exomoons in light curves has proven successful in the past, and similar network architectures achieved a validation accuracy of approximately 80-90 \% \citep{ExomoonCNN} (depending heavily on the noise level). I therefore also opted to use a CNN. The best-suited architecture for this application was determined experimentally by training and evaluating 40 different neural networks, each of which was given 1 to 5 convolutional layers and 1 – 8 dense layers. Additionally, in another experiment, the number of convolutional filters was gradually increased from 24 to 120 (where the step size was eight) in order to determine the best number of filters.
The final network architecture is a CNN with five convolutions, where each convolutional layer consists of a one-dimensional convolution, a batch normalization, a one-dimensional max-pooling layer and a dropout layer. The number of filters in the convolutional layers was set to 40 and a kernel size of 5 was used (this was also determined experimentally). The pooling layer uses a pooling size of 5. The dropout rate was set at 0.3 to prevent over-fitting.\footnote{I also tried other dropout rates, but with 0.3 I was able to achieve the best result, as no over-fitting occurs and the network also demonstrated effective learning.} These convolutional layers are followed by a flattening layer and three dense layers, whereby the first layer contains 64 neurons, the second 16 and the last layer 2 neurons, also forming the output layer. For a complete overview of the final network architecture, see the table \ref{tab:NetworkArchitecture}.

\subsection{Activation function}
Each layer requires a specific activation function. I decided to use the Rectified Linear Unit (ReLU). This function has become very popular in recent years:  
\begin{equation}
 \begin{array}{cc}
 f(x)=max(0,x)  & \text{(ReLU)} \\
 \end{array}
\end{equation}the use of the ReLU function reduces the error rate significantly better than, for example, the tangent hyperbolicus, as shown in \cite{krizhevsky2012imagenet}
\begin{equation}
\begin{array}{cc}
tanh(x)=2\sigma(2x)-1 & \text{(tanh)}
\end{array}
\end{equation}
with \(\sigma(x) = \frac{1}{1+e^{-2x}}\). The output layer activation function is a softmax function that converts the output of the hidden layers into a probability for the classification task. The sum of each probability of the classes is 1.

\subsection{Training and validation}
The Stochastic Gradient Descent Optimizer (SGD) is used to train the neural network. The learning rate of this optimizer was set using an exponential decay scheduler with an initial learning rate of 1E-2. The neural network is trained several times so as to obtain optimum training results: it has been found that results are very dependent on the initial sorting of the training data set. Each network is trained for 20 epochs; at the end of each epoch the model is saved and the validation accuracy is determined.\\
The light curves are loaded in groups of ten containing five noisy light curves with moon and five noisy light curves without moon. These light curves are divided into 12 parts, each with a length of 5,000 simulated photometric measurements. In this way, the size of the input layer can be reduced and the network is no longer too prone to over-fitting, as would be the case with a larger network. Due to a maximum planetary period of 60 days, each of these samples contains a transit signal. In addition, the data is normalized, which means that the maximum value in this data is 1 and the minimum value is 0. This makes the transit signals of the moon and the planet much more recognizable, as experiments without this normalization showed.\\
The accuracy and the loss function for training and for the validation run are determined for each epoch, where the accuracy is given by \cite{ExomoonCNN}: 
\begin{equation}
Acc=\frac{TP+TN}{TP+FP+TN+FN}
\end{equation}
 here, \(TP\)  is the number of light curves correctly classified as positive, \(FP\)  is the number of light curves incorrectly classified as positive, \(TN\) is the number of correct negatives and \(FN\) is the number of incorrect negatives. Although normally a binary or categorical cross-entropy is used as the loss function for these classification tasks, I used mean square error (MSE) since this enabled me to further improve the performance of the network.
 \subsection{Further development of the CNN}\label{WeiterentwicklungCNN}
\subsubsection{Architecture, loss function and data used}
Since it was not possible to apply the neural network effectively to real, i.e. non-simulated, light curves as described above, I developed another neural network. In order to ensure comparability of the different networks, I kept the architecture of the network the same as described above in \ref{architecture}. However, it turned out that for this network the MSE did not produce good results. For this reason, after several experiments, I used binary cross-entropy for this network, which is often used in classification tasks.
Another change made was in the size of the neural network. The input layer had a size of only 100 neurons. I used 100 neurons so that a section of the entire light curve of a star of about 2 days can be given as the sole input. These sections of the light curves contained precisely the transits of the moons and the planet, or only those of the planet. It is of course important for the exomoon signals to be located within this time frame, too; if they were not there, the network would not be able to detect them. The restriction by the Hill radius and the resulting maximum separation of the lunar transit from the planetary transit by the Hill time ensures that the lunar transit is within the selected section.

\subsection{Test method}
In order to be able to assess how well this neural network functions in practical use, I first tested it using the simulated data with which it had previously been tested during training. I plotted about 5,000 randomly selected transits from this data set and checked whether transits of an exomoon are visible in these plots. The trained CNN was also used to perform the same task on the same data. In addition, I selected the transits of the exoplanet HAT-P-7 b from the Kepler light curve. I received 618 transits in this way and ran them through the CNN. It is important to note that no exomoon signatures were visible in these transits. As such, I was able to verify the predictions of the neural network to see how well the network recognizes that there is no exomoon in the data. In addition, I selected the three known transits of the exoplanet Kepler-1625 b with the exomoon candidate Kepler-1625 b I from the publicly available data of the Kepler light curve and checked again whether an exomoon was detected. In this case, it would of course be ideal if the exomoon candidate could also be found by my CNN.
Furthermore, I conducted a sensitivity analysis for the ready-trained CNN. I simulated data sets in which all parameters were the same except for the radius of the moon. I varied the radius of the moon between \(0,1R_\oplus\) and \(6R_\oplus\). I injected the simulations into 20 different sections of the integration light curves so that 100 simulations were always inserted in the same section of an integration light curve. The noise in these light curves was about 50-100 ppm, thereby roughly corresponding to the noise of PLATO targets up to 11 mag.

\section{Results} \label{sectionResults}
\begin{figure}
\centering
 \includegraphics[width=8cm]{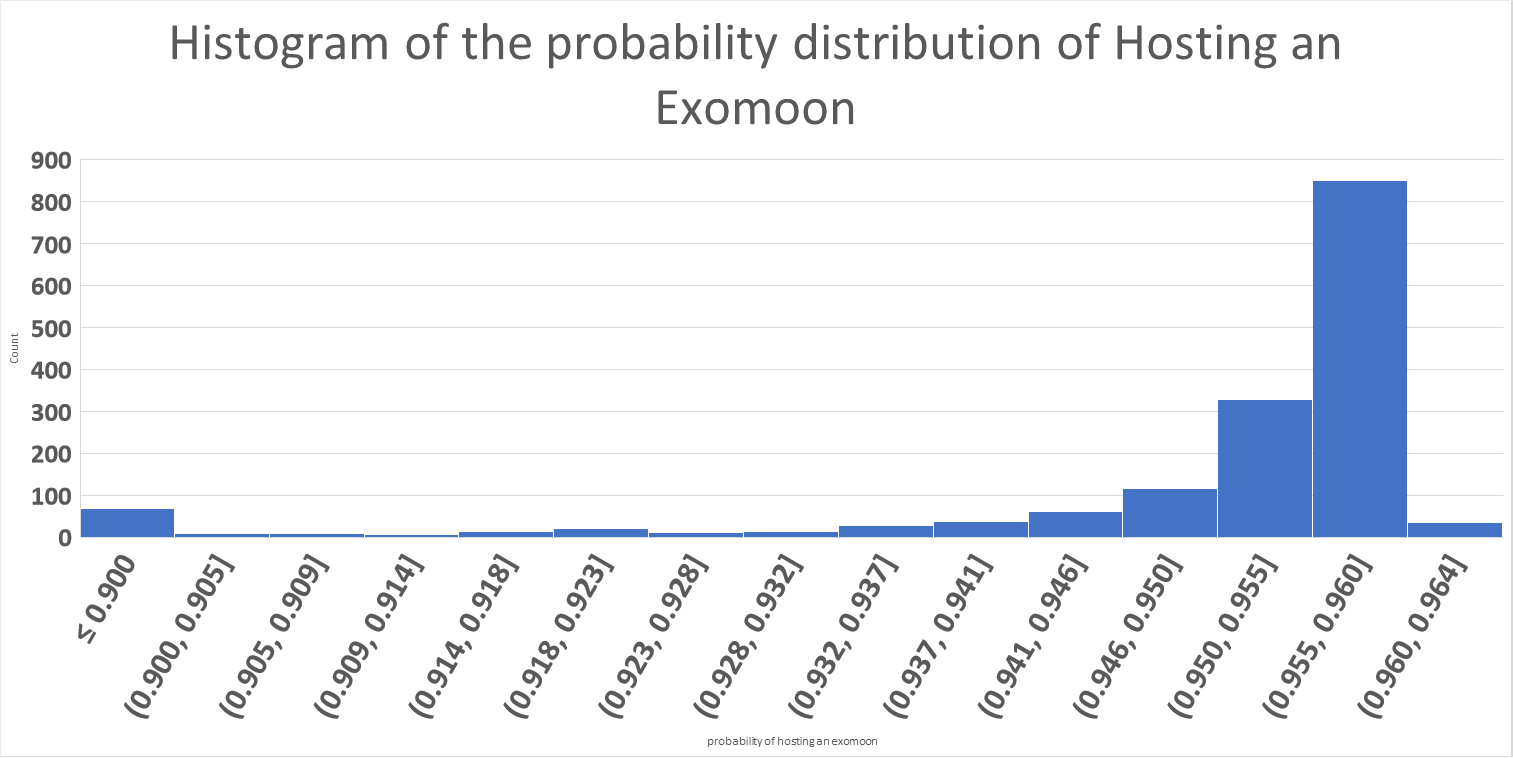}
 \caption{First CNN: This histogram shows that the CNN trained with the whole light curves almost exclusively predicts very high probabilities, which could potentially be due to a not yet perfect representation of stellar and instrumental noise in the network, for example.}
 \label{fig:PredictedProbabilitiesOverview}
\end{figure}
\begin{figure}
\includegraphics[width=8.8cm]{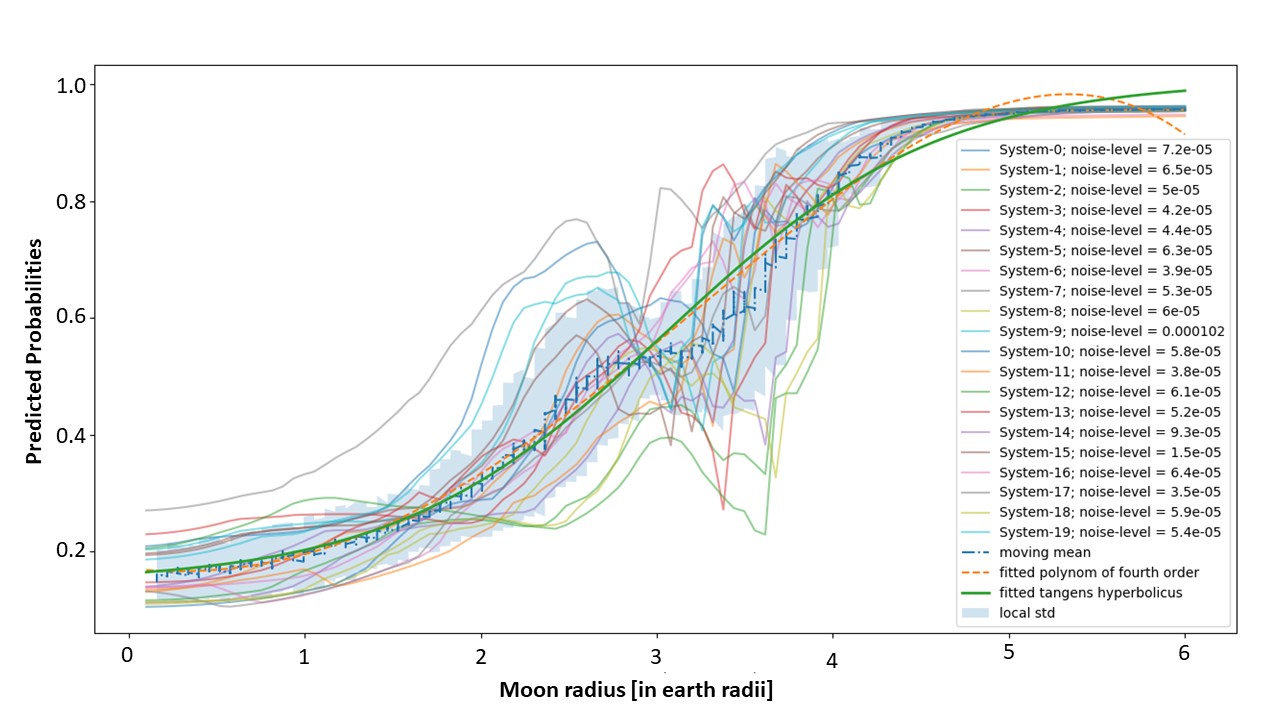}
\caption{Advanced CNN: In this diagram I show the results of the sensitivity analysis. The first 20 graphs represent the results for 2000 sample transits with each containing one moon signature (100 different sized moons per light curve). The bold green line represents a fitted tangens hyperbolicus, the orange dashed a fourth order polynomial and the blue dashed the moving mean. Here one can see that the larger the radius of the exomoon, the more clearly it is detected by the neural network. This is the postulated and desired response of the network, since a deeper, lunar transit naturally means that there is a greater probability of an exomoon.}
\label{fig:sensitivity}
\end{figure}
\subsection{General results}
Trying out different network architectures resulted in the network architecture described above (see \ref {architecture} and see table \ref{tab:NetworkArchitecture}). This network architecture uses 40 convolutional filters for one layer, resulting in a validation accuracy of about 89.0 per cent. Using a smaller number of filters reduced validation accuracy; when the number of filters was increased, this validation accuracy was still similar to that obtained by using 40 convolutional filters. In addition, it was possible to observe that the preprocessing of the data has an enormous influence on the network’s ability to distinguish between the two classes in the first place. Without normalizing the photometric measurements in a range between 0 and 1, the network was not able to distinguish between light curves that contained a moon and those in which no moon was present.\\
Training the neural network several times with the same data set was seen to result in significantly different results, possibly due to the random shuffling of the samples at the beginning of each epoch. The comparison of validation and training accuracy shows that 20 epochs is a good compromise between still (gradually) improving training accuracy while also preventing over-fitting, which would have been the case if the validation accuracy had fallen well below the level of the training accuracy.\\
Experiments with simpler data sets, where only t0\footnote{Here the time of the first planetary transit is given as t0.} was varied, had a higher validation accuracy of about 99.99 per cent (and training accuracy was also very high). Although this model exhibited a high level of accuracy, it cannot be used in practice because it is not able to make reliable statements in cases where other parameters vary, such as the moon-star radius ratio, since it has not been trained with such data.
\subsection{Results for the first CNN}
Applying the first model to the original Kepler data resulted in very high probabilities of an exomoon for each exoplanet (with the exception of some outliers) as it can be seen in Fig. \ref{fig:PredictedProbabilitiesOverview}. It can be assumed that this does not reflect the reality because the model should look for certain patterns but these patterns were not visible when controlling these light curves manually.  This could be due to more instrumental and stellar noise, which has a more complex distribution than the (in the first attempt) induced, normally distributed noise. Consequently, the version of the neural network described above did not allow exomoons to be detected in real data (i.e. original Kepler data). 
\subsection{Results for the advanced CNN}
By developing the CNN described in \ref{WeiterentwicklungCNN}, I was able to detect exomoons in simulated data as well as in the combination of simulated\footnote{This combination is described in the lower paragraph of \ref{simulation}.} and original Kepler data (with low noise levels). Furthermore, I tested the CNN by using the light curves of two exoplanets. When I entered the available transits of HAT-P-7b as input to the neural network, it gave me a low probability for most transits that HAT-P-7b has an exomoon. This fits well with the “manual” verification of the transits. By contrast, the transits of Kepler-1625 b generated a high probability that an exomoon exists there. This is also a very good result, since Kepler-1625 b is thought to have an exomoon. But at the same time, it can not be surely excluded whether this was due to the potential exomoons signature or due to a higher noise level.\\
 It was also possible to show with my sensitivity analysis that I can effectively detect exomoons from a size of approx. \(2-3R_\oplus\) (see Fig. \ref{fig:sensitivity}).
 
 \section{Conclusions}
 \cite{ExomoonCNN} already trained a network on simulated data in the search for exomoons; I used a similar network architecture but determined more hyper-parameters of the CNN experimentally. Furthermore, I simulated different exomoons and planets (i. e. which orbit the host star faster). Neural networks need to be trained with a huge amount of data; in my study, I used a combination of simulated synthetic light curves and observed high-quality Kepler light curves to enable the CNN to differentiate between exomoon signatures and  noise caused for example by stellar activity. Thus far, there was no exomoon candidate confirmed so the simulation of synthetic light curves was necessarily needed even if some candidates would be confirmed (this would not be enough training data). The main findings in this research project can be summarized as the follows:
\begin{enumerate}
    \item Convolutional Neural Networks seem to be not only able to find exomoons signatures (i. e. transits) in simulated data as shown by \cite{ExomoonCNN}, but also might be able to find bigger exomoons in real observations.
    \item To get realistic results when dealing with real observations, the network should be trained by using synthetic light curves to ensure that in the training set are "confirmed" exomoon transits and not only possible artifacts of noise. Adding Kepler light curves that show no exoplanet transits and have low noise levels to these synthetic transits seem to support the networks ability of distinguishing between noise and exomoon transits.
    \item The network architecture used by \cite{ExomoonCNN} could be reconfirmed and some further hyper-parameters could be determined.
    \item It has been shown that using clippings of the whole light curve which contains the planetary and lunar transits (200 photometric measurements per transit) instead of the whole light curve as the input of the neural network helps to prevent it from identifying noise as a exomoon signature. 
\end{enumerate}
This has the potential to significantly accelerate data processing for the PLATO mission due to be launched in 2026.

\section*{Acknowledgments}
This project was supervised by Dr. Philipp Eigmueller (German Aerospace Center, DLR). I got further support from Dr. Rolf Kuiper (Institute of Theoretical Astrophysics, University of Heidelberg) and Dr. Alexis Smith (German Aerospace Center, DLR). Thanks for the support. This paper includes data collected by the Kepler mission and obtained from the MAST data archive at the Space Telescope Science Institute (STScI). Funding for the Kepler mission is provided by the NASA Science Mission Directorate. STScI is operated by the Association of Universities for Research in Astronomy, Inc., under NASA contract NAS 5–26555.

\bibliography{main.bib}

\begin{appendix}
\onecolumn
\section{Further parameters used to simulate the synthetic light curves}\label{otherParameters}

\begin{table}[!hbt]
\caption{These star parameters were used in addition to Table \ref{tab:parametersStellar} to simulate the light curves for the training and validation dataset.}
\centering
\begin{tabular}{cc}
\hline  \hline
\textit{type}                          & \textit{value} \\ \hline
Orbital period in days                 & 0 d            \\
Orbital inclination in degrees         & 90°            \\
Orbital eccentricity                   & 0              \\
Longitude of pericenter in degrees     & 0°             \\
Longitude of ascending node in degrees & 0°             \\
Time of primary eclipse in days        & 0 d            \\
effective temperature in Kelvin        & 5577           \\
type of limb-darkening                 & quadratic      \\ \hline
\end{tabular}
\end{table}
\begin{table}[!hbt]
\centering
\caption{The planetary and lunar parameters shown in this table have an additional influence on the simulated light curve but were not varied during the simulation of the different data sets (see Table \ref{tab:parametersMoon} and Table \ref{tab:parametersPlanet} for the other parameters).}
\begin{tabular}{cc}
\hline  \hline
\textit{type}                          & \textit{value}    \\ \hline
Longitude of ascending node in degrees & 0°                \\
Longitude of pericenter in degrees     & 0°                \\
albedo                                 & 0.3               \\
Nightside temperature in Kelvin        & 40                \\
limb darkening type                    & no limb darkening \\ \hline
\end{tabular}
\end{table}

\pagebreak
\section{Network architecture}
\begin{table}[!hbt]
\caption{This table shows the CNN's experimentally-determined architecture which offered the best match.}
\centering
\begin{tabular}{|c|c|c|c|}
\hline
\textit{type}       & \multicolumn{3}{c|}{\textit{hyperparameters}}                     \\ \hline
1D Convolution      & number of filters = 40 & kernel-size = 5      & activation = ReLU \\ \hline
Batch-Normalization &                        &                      &                   \\ \hline
1D MaxPooling       & pool-size = 5          &                      &                   \\ \hline
Dropout             & Dropout-rate = 0.3     &                      &                   \\ \hline
1D Convolution      & number of filters = 40 & kernel-size = 5      & activation = ReLU \\ \hline
Batch-Normalization &                        &                      &                   \\ \hline
1D MaxPooling       & pool-size = 5          &                      &                   \\ \hline
Dropout             & Dropout-rate = 0.3     &                      &                   \\ \hline
1D Convolution      & number of filters = 40 & kernel-size = 5      & activation = ReLU \\ \hline
Batch-Normalization &                        &                      &                   \\ \hline
1D MaxPooling       & pool-size = 5          &                      &                   \\ \hline
Dropout             & Dropout-rate = 0.3     &                      &                   \\ \hline
1D Convolution      & number of filters = 40 & kernel-size = 5      & activation = ReLU \\ \hline
Batch-Normalization &                        &                      &                   \\ \hline
1D MaxPooling       & pool-size = 5          &                      &                   \\ \hline
Dropout             & Dropout-rate = 0.3     &                      &                   \\ \hline
1D Convolution      & number of filters = 40 & kernel-size = 5      & activation = ReLU \\ \hline
Batch-Normalization &                        &                      &                   \\ \hline
1D MaxPooling       & pool-size = 5          &                      &                   \\ \hline
Dropout             & Dropout-rate = 0.3     &                      &                   \\ \hline
Flatten             &                        &                      &                   \\ \hline
Dense               & number of neurons = 64 & activation = ReLU    &                   \\ \hline
Dense               & number of neurons = 16 & activation = ReLU    &                   \\ \hline
Dense               & number of neurons = 2  & activation = Softmax &                   \\ \hline
\end{tabular}
\label{tab:NetworkArchitecture}
\end{table}
\end{appendix}

\end{document}